\documentclass[oldversion]{aa}
\usepackage{graphicx}
\usepackage{txfonts}
\usepackage{natbib}
\usepackage{wrapfig}
\def\na1{Na{\sc I}}

\def\nh{cm$^{-2}$}
\def\fe{ergs\,cm$^{-2}$\,s$^{-1}$\,keV$^{-1}$}

\begin{document}

   \title{The isolated neutron star RBS1774 revisited\thanks{Based
   on observations obtained with XMM-Newton, an ESA 
   science mission with instruments and contributions directly funded by ESA
   Member States and NASA}\fnmsep\thanks{Based on data acquired using the
   Large Binocular Telescope (LBT). The LBT is an international collaboration
   among institutions in the US, Italy, and Germany. LBT Corporation partners
   are the University of Arizona, on behalf of the Arizona university system;
   Istituto Nazionale di Astrofisica, Italy; LBT Beteiligungsgesellschaft,
   Germany, representing the Max Planck Society, the Astrophysical Institute
   Potsdam, and Heidelberg University; Ohio State University; and the Research
   Corporation, on behalf of the University of Notre Dame, the University of
   Minnesota, and the University of Virginia.}} 
\subtitle{Revised XMM-Newton X-ray parameters and an optical counterpart
   from deep LBT-observations}
   \author{A. Schwope\inst{1}
     \and T. Erben\inst{2}
     \and J. Kohnert\inst{1}
     \and G. Lamer\inst{1}
     \and M. Steinmetz\inst{1} 
     \and K. Strassmeier\inst{1} 
     \and H. Zinnecker\inst{1} 
     \and J. Bechtold\inst{3}
     \and E. Diolaiti\inst{4}
     \and A. Fontana\inst{5}
     \and S. Gallozzi\inst{5}
     \and E. Giallongo\inst{5}
     \and R. Ragazzoni\inst{6}
     \and C. De Santis\inst{5}
     \and V. Testa\inst{5}
}

   \institute{Astrophysikalisches Institut Potsdam,
     An der Sternwarte 16, 14482 Potsdam, Germany\\
     \email{aschwope@aip.de}
     \and
     Argelander-Institut f\"ur Astronomie (AIfA), University of Bonn, 
     Auf dem H\"ugel 71, 53121 Bonn, Germany
     \and
     Steward Observatory, University of Arizona, Tucson, AZ 85721, USA
     \and
     INAF, Osservatorio Astronomico di Bologna, via Ranzani 1, 
     40127 Bologna, Italy
     \and
     INAF, Osservatorio Astronomico di Roma,
     Via di Frascati 33, I-00040 Monteporzio, Italy
     \and
     INAF, Osservatorio di Padova, vicolo dell'Osservatorio 5, I-35122 Padova,
     Italy 
      }

   \date{Received ; accepted }
\abstract{We report optical B-band observations with the Large Binocular
Telescope LBT of the isolated neutron star RBS1774. The stacked
image with total exposure $2\fh5$ 
reveals a candidate optical counterpart at $m_B = 26.96 \pm 0.20$ at
position $\alpha {\rm (2000)} = 21^h43^m03\fs40$, 
$\delta{\rm (2000)} = +06\degr54^\prime17\farcs5$, within the joint Chandra and
XMM-Newton error circles. We analyse archival XMM-Newton observations and 
derive revised spectral and positional parameters. 
The predicted optical 
flux from the extrapolated X-ray spectrum is likely twice as high as reported before. 
The measured optical flux exceeds the extrapolated
  X-ray spectral flux by a factor $\sim40$ (15 -- 60 at 1$\sigma$ confidence). 
We interpret our detection and the
  spectral energy distribution as further evidence of a 
  temperature structure over the neutron star's surface and present a pure
  thermal model reflecting both the SED and the pulsed fraction of the light
  curve. 
}

\maketitle

\section{Introduction}
RBS1774 was one of the few X-ray bright, high-latitude
sources found in the ROSAT all-sky survey (RASS) which remained undetected at
other wavelengths despite considerable effort \citep{schwopeetal00}. Based on
the large X-ray to optical flux ratio implied by the non-detection at optical
wavelengths, $m_R > 23$, $f_{\rm X}/f_{\rm opt} > 10^3$,
\citet{zampierietal01} tentatively identified the 
object as Isolated Neutron Star (INS). 
The ROSAT X-ray spectrum was fitted with a
hot blackbody, $kT \sim 90$\,eV, slightly absorbed by cold interstellar
matter with a total column density of $N_{\rm H} \sim 5 \times 10^{20}$\,\nh.
First observations with XMM-Newton by \citet{zaneetal05} revealed
pulsations with a period of 9.437\,s with a pulsed fraction of about 4\%. They
also revealed a spectral feature at 0.7\,keV. 
These findings added to the meanwhile accepted picture of XDINS
(X-ray Dim INS) being a class of middle-aged, $\sim 10^6$\,yrs, cooling NS
with strong magnetic fields, $B > 10^{13}$\,G \citep[see][for a recent review
of the class]{haberl07}.

Optical counterparts of four XDINS were detected with the HST, VLT, NTT, and
the SUBARU telescopes in the magnitude range 25.7\dots28.6. In all cases the
measured optical flux exceeded the extrapolated X-ray flux, typically assumed
to be blackbody-like, by considerable factors. 
The excess flux was qualitatively  
explained as originating from cooler areas of the structured NS surface not
detected in X-rays, see \citet{schwopeetal05} or \citet{zaneturolla06} for
light 
curve and SED modeling of XDINS. Furthermore, 
optical observations are important to search for
parallaxes and proper motions \citep{motchetal05,motchetal09}
and thus constrain the likely birth place of the stars.

Despite considerable effort through multi-site, multi-band observations,
RBS1774 escaped an optical detection in the past
\citep{zampierietal01,cropperetal07,locurtoetal07}. The latest compilation of
optical upper limits from multi-band observations 
is given in \citet{reaetal07} with a most sensitive limit 
$m_V > 25.5$. 

However, \citet{malofeevetal06} report a detection at radio
wavelengths at 111.23\,MHz at a brightness of $60\pm25$\,mJy which, if
confirmed, gives evidence for a non-thermal component in this system. 

The optical non-detection gave reason to search for an
optical counterpart with the newly opened Large Binocular Telescope (LBT).
The object was chosen as an appropriate target in Science Demonstration
Time (SDT) in October 2007. The observations revealed a positive
detection of a likely optical counterpart. 

While still analysing our LBT-data a detection of the likely optical
counterpart of RBS1774 with VLT/FORS was reported by 
\citet{zaneetal08}. Hence, the current paper presents a comparison of the
optical 
results and also presents improved X-ray spectral parameters compared to
the previous analysis of the same data by \citep{zaneetal05, cropperetal07,
  reaetal07}.  
\section{Optical observations} 
RBS1774 was observed with LBC-B, the blue arm of the Large Binocular Camera
\citep{giallongoetal08}, 
during five nights in October 2007 under mostly photometric conditions. 
The number of frames taken together with individual exposure times are
collected in Tab.~\ref{t:obs}. 
Small positional offsets between individual exposures using a
variety of dither patterns ensured a continuous spatial coverage without chip
gaps. 
The data were reduced with the GaDoDS pipeline
\citep{erbenetal05}, which includes basic CCD calibrations (dark, bias,
flatfield correction) as well as superflat correction, astrometric and
photometric calibration, and creation of a stack of all images.

\begin{table}[t]
\caption{Summary of LBT observations of the field of RBS1774. All exposures
  were taken through a Bessel B-filter. The last column gives the nightly
  photometric zeropoint.}
\begin{tabular}{lrrrc}
\hline
Date & frames$\times$ exposure & Exposure tot & $b_0$\\
YYYYMMDD&&(s) \\
\hline
20071012 & $10\times300$ & 3000 & 27.68\\
20071014 & $10\times300$ & 3000 & 27.61\\
20071015 & $ 5\times300$ & 1500 & 27.61\\
20071016 & $16\times300$,$10\times150$ & 6300 & --\\
20071018 & $ 2\times150$ &  300 & 27.62\\
\hline
\label{t:obs}
\end{tabular}
\end{table}
 
A global astrometric solution was achieved using the GSC2-2. 
and was based on 414 (GSC) high S/N detections revealing an 
external astrometric
accuracy $\sigma = 0\farcs165$ in both spatial directions. 
  
Photometric zeropoints were determined for each night (except Oct.~16) 
based on observations of Stetson's standard fields. We did not re-calibrate
the extinction and 
color-terms but used the values determined by the instrument team during
commissioning of the 
instrument\footnote{http://lbc.oa-roma.inaf.it/commissioning/standards.html}.
The nightly zeropoints $b_0$ are also listed in Tab.~\ref{t:obs}.  

The seeing varied between 0\farcs76 and 1\farcs31 while exposing on RBS1774. 
We created two versions of a stacked image, one with the best spatial 
resolution using only images with a seeing better than 1\farcs0 (25
images, total exposure 5700\,s, seeing FWHM $\sim$0\farcs88), 
and a second image with fainter limiting magnitude using all images 
with negligible ellipticity of stellar profiles (35 frames, total exposure
9010\,s, FWHM seeing $0\farcs93$).  

Source catalogues of the fully reduced B-band 
images were constructed using {\tt SExtractor} \citep{bertinarnouts96}. 
For brightness determination of faint pointlike objects we use aperture
magnitudes with a diameter of 8 pixels (corresponding to 1.8 arcsec). 
The raw magitudes were corrected 
for light loss through the rather small aperture. The magnitude correction was
determined by a growth curve analysis of a stellar sequence in the magnitude
range 20--23 and was found to be 0.22\,mag for the chosen aperture. All
magnitudes quoted in this paper are in the Vega system.

\begin{figure}
\centering
\resizebox{\hsize}{!}{
\includegraphics[bbllx=70pt,bburx=552pt,bblly=48pt,bbury=705pt,angle=-90,clip=]{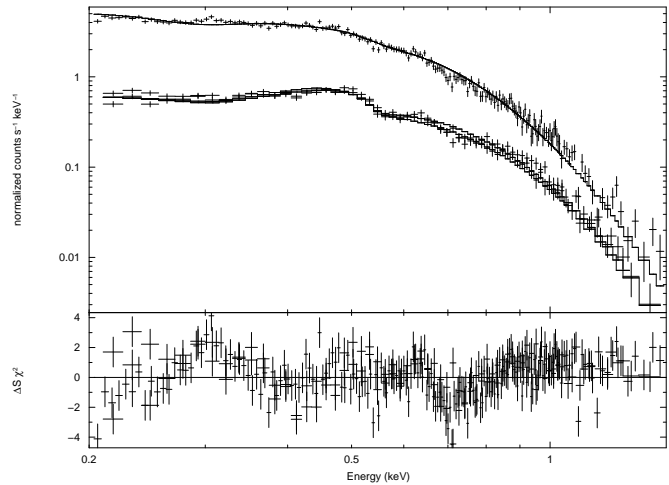}
}
\caption{Combined XMM-Newton EPIC-pn and EPIC-MOS spectra of RBS1774, fitted
  with a Planckian absorbed by cold interstellar matter (model (1) in
  Tab.~\ref{t:xfit}).
\label{f:xfit1}
}
\end{figure}
\begin{figure}
\resizebox{\hsize}{!}{
\includegraphics[bbllx=70pt,bburx=552pt,bblly=48pt,bbury=705pt,angle=-90,clip=]{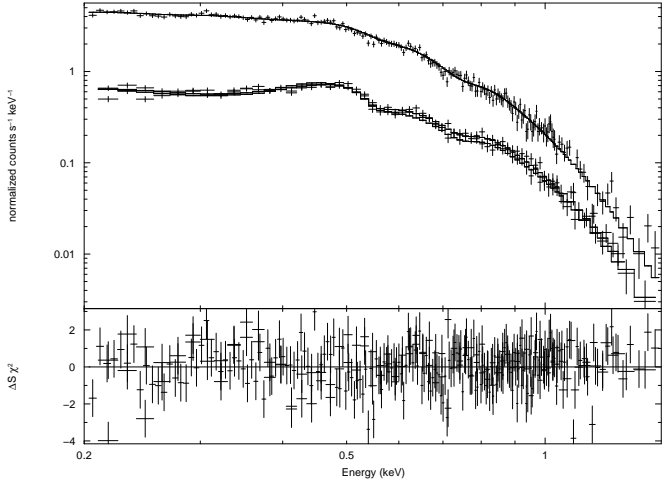}
}
\caption{Same data as in Fig.~\ref{f:xfit1}, fitted
  with a Planckian and superposed Gaussians absorbed by cold interstellar
  matter (model (3) in Tab.~\ref{t:xfit}). 
\label{f:xfit2}
}
\end{figure}

\begin{figure}
\resizebox{\hsize}{!}{
\includegraphics[bbllx=70pt,bburx=552pt,bblly=48pt,bbury=705pt,angle=-90,clip=]{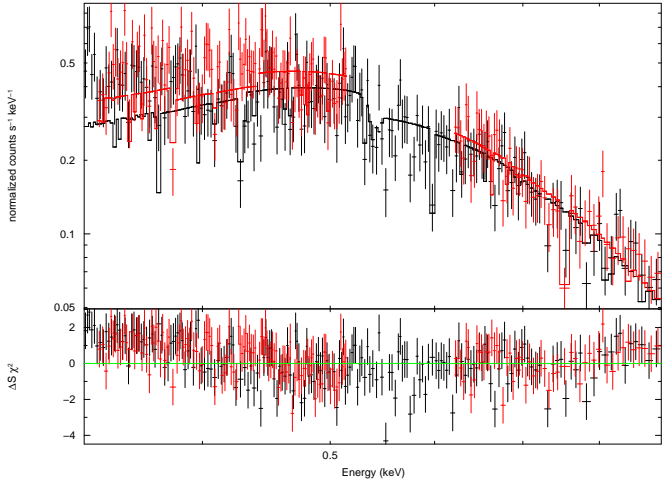}
}
\caption{RGS-spectra of RBS1774, fitted with a Planckian absorbed by cold
  interstellar matter (see text for details). The data gap between 0.5 
and 0.6 keV is due to the dead CCD in RGS2.
\label{f:xfit3}
}
\end{figure}

\begin{figure}
\resizebox{\hsize}{!}{
\includegraphics[bbllx=70pt,bburx=552pt,bblly=48pt,bbury=705pt,angle=-90,clip=]{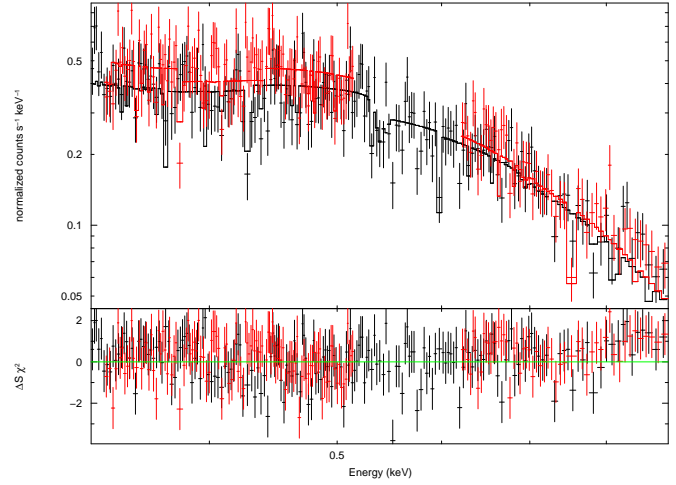}
}
\caption{RGS-spectra of RBS1774, fitted with two Planckians absorbed by cold
  interstellar matter. The temperatures were fixed at $k_{\rm B}T_{\rm spot} =
  104$\,keV and $k_{\rm B}T_{\rm star} = 40$\,eV, respectively (model (6) in
  Tab.~\ref{t:xfit}).
\label{f:rgs_fin}
}
\end{figure}

\section{XMM-Newton X-ray observations}
There is one published XMM-Newton observation of RBS1774 which was performed
in 2004. It was
analysed and discussed by \citet{zaneetal05} and \citet{cropperetal07} who
give all the neccessary details of the observations. Main
results of those studies were the detection of the spin frequency of 9.437\,s
and of a spectral feature at 0.7 keV which could be modeled as an absorption
edge or line superposing the blackbody-like continuum. 

\subsection{Astrometric analysis}
Giving the well matching large fields of view of the LBC and the XMM-Newton
EPIC cameras we attempted to improve the X-ray positional accuracy. We make use
of the X-ray source list of the field as provided in the XMM-Newton Science
Archive (XSA) and in the 2XMM catalogue which contains 43 sources including
the target. The X-ray 
coordinates of RBS1774 as given in 2XMM are $\alpha(2000) = 21^{\rm
  h} 43^{\rm m} 03\fs22$, $\delta(2000) = +06^{\rm h} 54\arcmin 18\farcs0$) with an
uncertainty of $\sigma = \sqrt{\sigma_{\rm stat}^2 + \sigma_{\rm sys}^2} =
\sqrt{0.07^2+1.0^2} \simeq 1\farcs0$. 

The optical and X-ray source catalogues were matched using
SAS-task {\tt eposcorr} to remove remaining systematic uncertainties of the X-ray
source positions due to imperfections of the spacecraft attitude. 
{\tt Eposcorr} reported 23 X-ray/optical matches
and revealed positional offsets $\Delta\alpha=3\farcs0\pm0\farcs6$, $\Delta\delta
=0\farcs2\pm0\farcs5$, and a rotation correction of $30\arcsec\pm170\arcsec$. The
corrected XMM-Newton position of RBS1774 is $\alpha = 325\fdg7643 = 21^{\rm
  h} 43^{\rm m} 03\fs4$, $\delta = 6\fdg9050 = +06^{\rm h} 54\arcmin
17\farcs9$) with a 1$\sigma$ positional uncertainty of only 0\farcs7.

The X-ray source position of RBS1774 is shifted by $2\farcs7$ with 
respect to the entry in the 2XMM source catalogue. 
The accuracy of the XMM-Newton X-ray position was overestimated in 2XMM, the
revised position is more reliable thanks to the available optical
reference positions from the wide-field LBC-observations. 
We note in particular that the Chandra position (we use the revised
position by Zane et al.~2008)
and the revised XMM-Newton positions match excellently. 

\begin{table*}[t]
\caption{X-ray spectral parameters of RBS1774 for combined fits to
  EPIC-pn and EPIC-MOS in the energy interval $0.2-1.5$\,keV. 
  The errors quoted correspond to 90\% confidence intervals. 
  Lines (4) and (5) represent combined fits to EPIC and RGS using same models
  as in lines (1) and (2). Line (6) lists the finally accepted
  two-component blackbody fit to the RGS-data alone with fixed ratio of
  normalization parameters.
\label{t:xfit}}
\begin{tabular}{rlllllll}
\hline
No. & Model & $N_{\rm H}$ & $kT_{\rm bb}^\infty$ & $E_{\rm line}$ & $\sigma_{\rm 
  line}$ & $\tau_{\rm line}$/Norm$_{\rm line}$ & $\chi^2_\nu$ (d.o.f.) \\
& & ($10^{20}$\,cm$^{-2}$) & (eV) & (eV) & (eV) & (\# cm$^{-2}$ s$^{-1}$) & \\
\hline
(1) & bbody & $1.85^{+0.17}_{-0.17}$  & $103.5\pm0.8$ &
& & & 
1.81(311) \\
(2) & bbody*gabs & $1.84^{+0.20}_{-0.17}$ & $105.1\pm0.9$ & 
$0.731^{+0.008}_{-0.013}$ & 
$27^{+16}_{-4}$ & 
$6.5^{+1.2}_{-1.0}$ & 
1.50(308)\\
(3) & bbody*gabs & $3.92^{+0.75}_{-0.50}$ & $102.7\pm0.8$ & 
$0.733^{+0.008}_{-0.012}$ & 
$34^{+11}_{-11}$ & 
$7.3^{+1.3}_{-1.0}$ & 
\\
& \phantom{bbody*gabs}+gaussian& & & 
$0.31$ & 
$0.03$ & 
$1.7^{+0.8}_{-0.5} \times 10^{-3}$ &
1.23(307)
\\
(4) & bbody & 
$1.62^{+0.16}_{-0.14}$ & 
$103.9\pm0.9$ & 
&
&
&
1.55(661)
\\
(5) & bbody*gabs & 
$1.64^{+0.17}_{-0.15}$ & 
$105.2\pm0.1$ & 
$0.729^{+0.004}_{-0.007}$ & 
$23^{+20}_{-19}$ & 
$4.8^{+1.7}_{-1.6}$ & 
1.42(658)
\\
(6) & bbody$+$bbody & 
$1.4^{+0.4}_{-0.4}$ & 
$104$! & 
\\
& & 
&
$40$! & 
&&& 1.05(349)
\\
\hline
\end{tabular}
\end{table*}

\subsection{Spectral analysis}
X-ray spectra of the three EPIC X-ray cameras were extracted
to estimate the contribution of the blackbody-like soft X-ray
component of RBS1774 to the optical B-band.
We re-processed the archival XMM-Newton data (OBSID 0201150101) with 
the Science Analysis Software (SAS) version 7.1. 
Subsequent X-ray spectral analysis
was performed with XSPEC V12.3.0. The X-ray spectra were grouped in spectral
bins containing at least 20 photons to allow the application of $\chi^2$
minimisation. 
The spectral analysis was limited to energies between 0.2 and
1.5\,keV. A fit with a pure blackbody
component absorbed by cold interstellar matter ({\tt phabs * bbody} in XSPEC
terms) gave a bad representation of the data and revealed large systematic
residuals (reduced $\chi^2_{\rm} = 1.81$ for 311 degrees of freedom, see
Fig.~\ref{f:xfit1} for a graph of the fit and the residuals). 
Compared to \citet{zaneetal05} and \citet{cropperetal07} we found the spectral
parameters slightly changed. The amount of
interstellar matter implied by the pure blackbody fit was found to be
smaller in our analysis, $N_{\rm H} = (1.85\pm0.17)\times 10^{20}$\,cm$^{-2}$
vs.~$N_{\rm H} = 3.65^{+0.16}_{-0.13} \times 10^{20}$\,cm$^{-2}$, while the temperature was
found to be slightly higher, $kT_{\rm bb}^\infty = 103.5\pm0.8$\,eV
vs.~$101.4^{+0.5}_{-0.6}$\,eV (see Table~\ref{t:xfit} for a compilation of
spectral parameters). These differences might be due to the 
revised calibration of the soft X-ray spectral response of the
EPIC cameras. 

The feature at $\sim$0.7\,keV firstly reported by \citet{zaneetal05} is also
evident in our spectra. 
Including a Gaussian absorption line at $\sim$0.7\,keV
improved the fit ($\chi^2_\nu = 1.50$ for 308 degrees of freedom, see
Table~\ref{t:xfit}, line (2)) but still did not reveal a convincing
representation of the data. The best-fit width of this line of 27\,eV is
narrower than the CCD resolution at this energy of $\sim$40\,eV but still
consistent at 90\% confidence. 

Large residuals around 0.31\,keV in the EPIC-pn and one of the MOS
cameras are evident from inspection of Fig.~\ref{f:xfit1}. This feature,
described as an absorption line at $\sim$0.4\,keV, was noted and discussed
by \citet{cropperetal07}. 
A feature of possible similar nature was detected in EPIC-pn spectra of the
much 
brighter prototypical object RX\,J1856.4-3754 and classified as remaining
calibration problem by \citep{haberl07}. The same may apply to the current
case of RBS1774.

However, a feature at $0.3-0.4$\,keV influences our determination of the amount of
interstellar absorption and of the blackbody parameters. We thus included in
our fit a further spectral component, a Gaussian emission line centered at
0.31\,keV (model~(3) in Tab.~\ref{t:xfit}), 
to study the effect of modified spectral parameters on the
predicted flux in the B-band. 
We do not claim physical reality of an
emission line, but include it in the fit just to explore possible systematic
uncertainties of the $T_{\rm bb}$ and $N_{\rm H}$ determination. An emission
line raises $N_{\rm H}$ and hence raises the optical excess (see below) to a
maximum value.

We searched for independent information for both features by
extracting the spectra obtained with the RGS spectrographs onboard
XMM-Newton. Both RGS' operate in the wavelength range $5 - 35$\,\AA\ 
($0.33-2.5$\,keV), hence the
feature at 0.31\,keV is not fully covered. The EPIC and RGS spectra were
grouped and then firstly a combined fit was performed 
to the EPIC- and RGS-spectra (models (4) \& (5) in Tab.~\ref{t:xfit}) forcing
all parameters but the
normalisations for the two groups to be the same 
(reduced $\chi^2 = 1.42$ for 658 d.o.f.). Fixing then 
the best-fitting $N_{\rm H}$ and blackbody temperature from the combined fit
the RGS-spectra were fitted separately, one time with and another time 
without the Gaussian absorption component at 0.7\,keV. 
The reduced $\chi^2 = 1.33$ and
1.34 changed just marginally, hence the RGS spectra do not
support the existence of the 0.7\,keV feature. 
(see Fig.~\ref{f:xfit3} for a blackbody fit to the RGS data). 
We conclude that the nature of both features
at 0.31\,keV and at 0.73\,keV, respectively,
is uncertain, both might be real but both might be related to
calibration uncertainties of the CCDs and the RGS at those very 
soft X-ray energies. In their analysis, \citet{cropperetal07} reach similar
conclusions, they accept the feature at 0.7\,keV since it could be accomodated
in their RGS-fit depending on the assumed continuum, but regard the feature
at 0.4\,keV as unlikely. We agree that more data are necessary to finally
prove or disprove the existence of those features.
   
The fit to the RGS-data alone, however, leaves large residuals. 
These are smoothly increasing towards softer X-rays, suggestive of a further 
low-energy radiation component. We thus attempted a fit to the RGS-data with
the  
 combination of two blackbodies, a hot component with parameters almost fixed
 by EPIC and a cool component responsable for the RGS soft excess. In a first
 attempt, all 
 parameters were allowed to vary freely except the temperature of the hot
 component which was fixed at 104\,keV. An excellent fit was achieved with a
 reduced $\chi^2 = 0.98$ for 347 d.o.f.~and $kT_{\rm cool} = 33$\,eV. The
 absorption column density of this unconstrained fit, $N_{\rm H} = 6.7\times
 10^{20}$\,cm$^{-2}$ however, exceeds the galactic value. Fixing $N_{\rm H}$
 to $2.0\times 10^{20}$\,cm$^{-2}$ as implied by model (2), deteriorates the
 result slightly but still gives an acceptable fit,  $kT_{\rm cool} =
 30$\,eV at $\chi^2 = 1.03$ (348 d.o.f.). This result may be regarded as
 observational evidence for the presence of a cooler radiation component in
 RBS1774.

\begin{figure}
\resizebox{\hsize}{!}{
\includegraphics[bbllx=38pt,bblly=191pt,bburx=575pt,bbury=629pt,clip=]{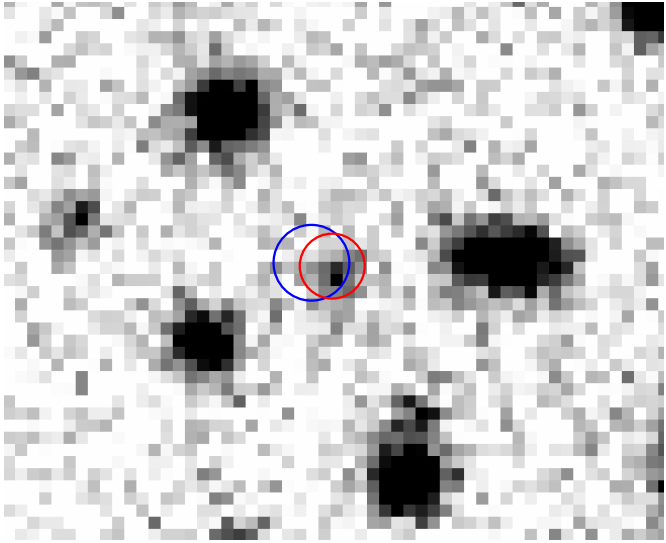}}
\caption{LBT B-band image of the region of RBS1774 with North at the top and East to
  the left (pixelsize $0\farcs225$). 
  Error circles illustrate the corrected Chandra (red) and XMM-Newton (blue) X-ray
  positions, respectively. 
\label{f:lbt}
}
\end{figure}

The amount of interstellar absorption, $N_{\rm H}$, was typically found to be
lower than the total galactic column density, $N_{\rm H, gal} = 4.83\times
10^{20}$\,cm$^{-2}$ in the direction of RBS1774 ($l^{II} = 62.66\degr$, $
b^{II} = -33.1\degr$). Only the inclusion of the additional absorption at 0.31\,keV
raised $N_{\rm H}$ close to its galactic value.
The unabsorbed flux in the B-band of the first 
three models listed in Tab.~\ref{t:xfit} lies between $F_{\rm E} = 2.8 \times
10^{-6}$\,\fe\ and $4.3 \times 10^{-6}$\,\fe\ at 90\% confidence (apparent
magnitude 30.9\dots30.4). 
Both, the minimum and the maximum value lie 
above the extrapolation of Zane's fit into the optical
\citep{zaneetal05,reaetal07}, $F_{\rm E} = 2.2 \times 10^{-6}$\,\fe\ ($m_B = 31.2$), 
the maximum flux estimated by us is  more than a factor 2 higher than the old
value. This is only partly due to the additional Gaussian, part of this effect
is possibly due to an improved calibration of the X-ray cameras.

\begin{figure}
\centering
\resizebox{\hsize}{!}{
\includegraphics[bbllx=55pt,bblly=34pt,bburx=540pt,bbury=532pt,angle=-90,clip=]{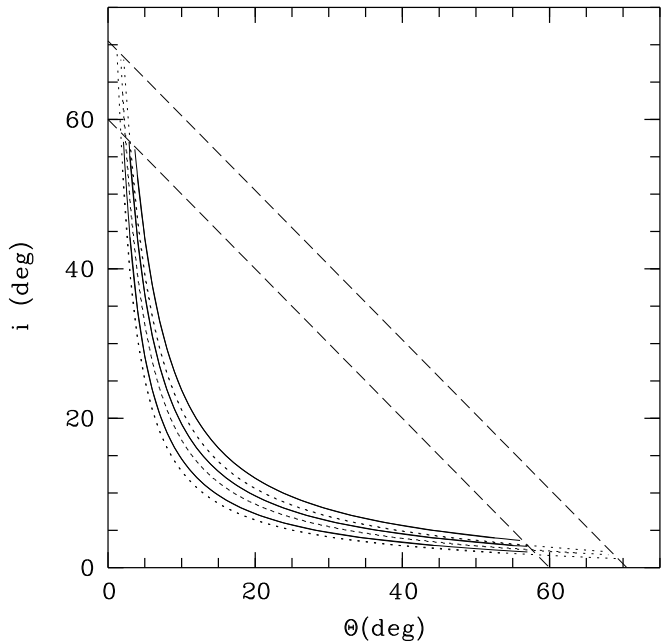}}
\caption{Spot locations in the $i-\theta$ plane asuming an emission radius $R
  = 3r_{\rm S}$ (sold line) and $4r_{\rm S}$ (dotted lines) and pulsed
  fractions of 3, 4, 5\%, respectively. The dashed lines mark the limit of
  class I light curves for the chosen compactness.
\label{f:i_t}
}
\end{figure}

\section{An optical counterpart to RBS1774 from LBC-imaging}
Fig.~\ref{f:lbt} shows a cut-out of the large LBT-image zoomed and centered on
the X-ray position of RBS1774. The revised Chandra \citep{zaneetal08} and
the revised XMM-Newton X-ray error circles are also shown. 
We confirm the detection of an optical counterpart to the X-ray source 
at a position of $\alpha {\rm (2000)} = 21^h43^m03\fs40$, 
$\delta{\rm (2000)} = +06\degr54^\prime17\farcs5$. 
It lies in the joint region of the error circles from both Chandra and
XMM-Newton. Positional offsets with respect to the X-ray positions are
$0\farcs21$ and $0\farcs55$, respectively. We cannot compare our optical 
position with that of \cite{zaneetal08}, since no absolute position is given
there, just a positional offset of $0\farcs2$ with respect to the revised
Chandra position, the same as ours. 
Their Fig.~1 and our Fig.~\ref{f:lbt} display
marginal apparent differences between the counterpart positions and the
Chandra error circles and we conclude that the optical source positions are
the same within the given errors.

The object detected by us has an apparent magnitude of  
$m_B = 26\fm96 \pm 0\fm15$ (1$\sigma$ statistical error). 
The color of the object is not known, a color correction
with a blue, blackbody-like object would change the apparent magnitude by
0.05\,mag which we include in the error budget. Hence, our final photometric
error is 0.2\,mag. 
Our measurement reveals a counterpart which is brighter by
0.4 mag compared to the detection by \cite{zaneetal08}. 
Both measurements have an uncertainty of 0.2\,mag. Hence, 
the brightness difference might be real.
If confirmed, the optical counterpart of RBS1774 would be the
first among its class displaying brightness variations. 

\begin{figure}
\centering
\resizebox{\hsize}{!}{
\includegraphics[bbllx=48pt,bblly=95pt,bburx=536pt,bbury=615pt,angle=-90,clip=]{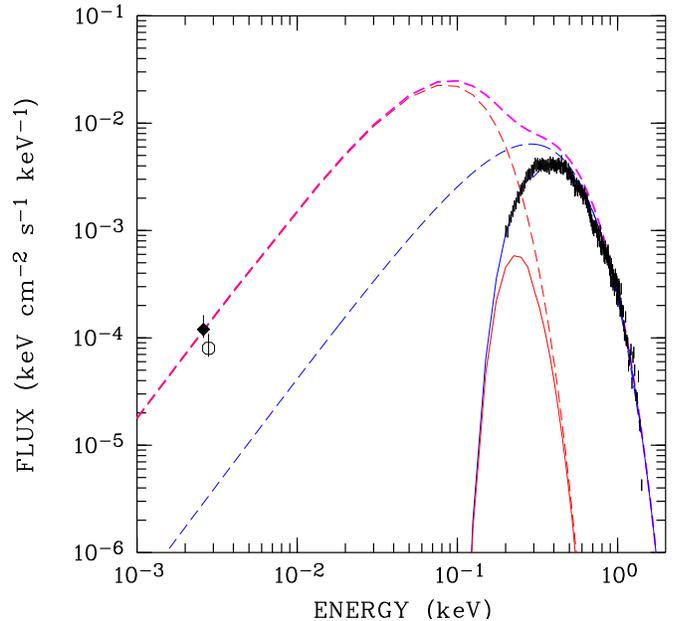}}
\caption{Spectral energy distribution of RBS1774. Shown are the unfolded X-ray
  spectrum in black, the absorption-corrected fit to the X-ray spectrum in
  blue (model 2 in Tab.~\ref{t:xfit}), the measured B-band fluxes corrected
  for interstellar absorption and an SED model involving a hot 104\,eV spot on a cool 40\,eV
  star (magenta, model (6) in Tab.~\ref{t:xfit}). 
  The absorbed and absorption-corrected contribution of the cool stellar
  surface is shown red.
\label{f:sed}
}
\end{figure}

\section{Discussion}
The amount of interstellar absorption as determined from the X-ray spectral
fits is in the range $N_{\rm H} = (1.8 - 3.6)\times10^{20}$\,cm$^{-2}$.
Assuming galactic dust with $R=3.1$ the absorption in the B-band is $\Delta
m_B = 0\fm13 - 0\fm27$, which we use to correct the magnitude found by us. 
Hence, the optical excess over the extrapolated flux from the X-ray 
spectral fits is
somewhere in the range between 15 and 60 (dependent on the assumed
brightness, the amount of interstellar absorption and the assumed X-ray
spectral parameters),  larger than for any other objects of its
class. Taken the parameters from X-ray spectral model (2) and $m_B = 26.96$,
$\Delta m_B = 0.13$ as face values, the optical excess is 40.

There are some lines of evidence for a non-homogeneous temperature
distribution over the star's surface, 
(a) a pulsed X-ray light curve,
(b) an optical excess, and 
(c) the likely presence of a second, cooler, component in the
  RGS-spectrum.
We firstly discuss the simplest model of
antipodal circular polar caps and a cooler stellar surface. 

Light bending and
light curves of antipolar caps were studied in the past by
\cite{pechenicketal83} and \cite{beloborodov02}. 
\citet{beloborodov02} classified types of light curves dependent on the viewing
geometry. According to this classification the light curve of RBS1774 is of
class I, with only one hot spot being persistently in view and the second
being never visible. The pulsed fraction gives  
an implicit equation between $i$ and $\theta$, the inclination of the rotation
axis and the latitude of the spot. The solution of this equation depends on
the assumed compactness of the star, i.e.~the radius of emission in units of
the Schwarzschild-radius $r_{\rm S}$. Solutions are shown in Fig.~\ref{f:i_t}
for 
$R_{\rm em} = 3 r_{\rm S}$ and  4$r_{\rm S}$, respectively and pulsed
fractions in the range 3, 4, 5\%, respectively. 
The X-ray light curve has a pulsed fraction of about 4\% \citep{zaneetal05}. 
The allowed ranges in $i$ and
$\theta$ go up to surprisingly large angles ($57\degr$ for $R_{\rm em}/r_{\rm
  S} = 3$, $68\degr$ for $R_{\rm   em}/r_{\rm S} = 4$), thus alleviating the 
conclusion of \citet{zaneetal08} that both angles are smaller than $20\degr$.
However, if one of the two angles takes its maximum value, the
  corresponding angle has to become very small. Such an extreme
  geometry seems unlikely and a geometry described by \cite{zaneetal08} more
  preferred, it seems to be, however, still compliant with the data.

The observed spectral energy distribution of RBS1774 from the optical to the
X-ray range is displayed in Fig.~\ref{f:sed}. Following \cite{schwopeetal05}
we synthesise the observed spectral energy distribution (SED) 
with the combination of a hot spot, $k_{\rm B} T_{\rm spot}$, and a cool
stellar surface, $k_{\rm B} T_{\rm star}$.

An indication of the possible surface temperature was   derived above,
 $k_{\rm B} T_{\rm star} \simeq 40$\,eV, which we assume in the following
 without loss of generality. 
A final fit to the RGS spectrum was performed with
 fixed temperatures, $k_{\rm B} T_{\rm spot} = 104$\, eV and $k_{\rm B} T_{\rm
 star}=40$\,eV, respectively, and a fixed ratio of the normalization
 parameters of the two blackbody components. A fixed ratio of the
 normalizations is required to reflect the observed optical excess. 
At the same time, it fixes the relative sizes of the spot and star radius 
(spot opening angle) for the chosen set of temperatures.
Free fit parameters were the column density
 and the normalization parameter of the
 hot component. The latter quantity in combination with the fixed ratio of
 normalization parameters encodes $R_{\rm ns, \infty}/d$. The fit was
 successful (reduced $\chi^2_\nu = 1.05$ for 348 d.o.f) 
and yielded a column density well below
 the galactic value, $N_{\rm H} = (1.4 \pm 0.4) \times10^{20}$\,cm$^{-2}$. The
 fit and the residuals are shown in Fig.~\ref{f:rgs_fin} (model parameters
 also listed in Tab.~\ref{t:xfit}, line (6)).

The same model applied to the three EPIC-spectra does not improve the fit
compared to model (2, single blackbody) and we conclude that either the
EPIC-spectrum is 
intrinsically more complex or that remaining calibration
uncertainties prevent final conclusions.

Given the simplicity of the model, the model SED shown in
Fig.~\ref{f:sed} is merely illustrative. It assumes a compactness $R_{\rm
  ns}/r_{\rm s} = 4$, $R_\infty = 13.8$\,km, a distance of 250\,pc, and a spot
opening angle of $6.5\degr$. The derived spot opening angle corresponds to a 
radius of 1.6\,km at 250\,pc which compares well with the estimate of
\citet{cropperetal07} who derive $r_X = 2$\,km at an assumed distance of
300\,pc. 

The chosen model successfully reflects the observed SED, but a
two-temperature atmosphere appears to be somehow artificial. 
Such a temperature distribution would likely require outside heating,
either from a magnetosphere or from accretion. 
Whether the most recent calculations of
neutron star crusts \citep{geppertetal06} could mimic such a temperature
distribution needs to be tested. Using our parameterization of the older
models by \citet{geppertetal04} we could not find a model reflecting the SED
with an optical excess of a factor 40. However, as we have shown, the optical
excess is uncertain and needs a more precise determination with better data.  

The distance of 250\,pc shall be regarded as lower limit to the star. Any
smoother and thus more realistic temperature distribution will yield a
larger distance.  

Whether the currently available data on RBS1774 are compliant with just
thermal radiation needs further investigation. 
\cite{zaneetal08} discuss alternative possibilities for the observed optical 
excess in RBS1774, powered either by magnetospheric radiation or by magnetic 
field decay, both leading to a non-thermal radiation component. The alternative
possibilities can be distinguished rather straightforwardedly by further
multi-band deep imaging.

\begin{acknowledgements}
The authors thank the LBT Science Demonstration Time (SDT) team for assembling
and executing the SDT program. We also thank the LBC team and the LBTO staff
for their kind assistance.

We thank an anonymous referee for helpful comments which improved the 
presentation of the paper. 

JK and TE were supported by the DFG priority program SPP-1177 ``Witnesses of
Cosmic History: Formation and evolution of black holes, galaxies and their
environment'' (projects Schw536/23-1 and ER327/2-2). GL was supported by the
Deutsches Zentrum f\"ur Luft- und Raumfahrt (DLR) GmbH under contract No.~50
OX 0201 and FKZ 50 QR 0802.
\end{acknowledgements}

\bibliographystyle{aa}
\bibliography{rbs1774_lbt}
\end{document}